\documentclass[sigconf]{acmart}

\usepackage{xspace}
\usepackage[capitalise]{cleveref}
\usepackage[colorinlistoftodos,prependcaption,textsize=tiny]{todonotes}

\graphicspath{{./}{./img/}}

\AtBeginDocument{%
	}

\copyrightyear{2024}
\acmYear{2024}
\setcopyright{acmlicensed}\acmConference[IDE '24]{2024 First IDE Workshop}{April 20, 2024}{Lisbon, Portugal}
\acmBooktitle{2024 First IDE Workshop (IDE '24), April 20, 2024, Lisbon, Portugal}
\acmDOI{10.1145/3643796.3648462}
\acmISBN{979-8-4007-0580-9/24/04}

\begin{document}
	
	\title{An IDE Plugin for Gamified Continuous Integration}

	\author{Philipp Straubinger}
	\affiliation{%
		\institution{University of Passau}
		\city{Passau}
		\country{Germany}
	}
	
	\author{Gordon Fraser}
	\affiliation{%
		\institution{University of Passau}
		\city{Passau}
		\country{Germany}
	}
	
	\renewcommand{\shortauthors}{Straubinger et al.}
	
	\newcommand{\toolname}{\textsc{Gamekins IntelliJ plugin}\xspace}
	\newcommand{\gamekins}{\textsc{\mbox{Gamekins}}\xspace}
	\newcommand{\Jenkins}{\textsc{Jenkins}\xspace}
	\newcommand{\IntelliGame}{\textsc{\mbox{IntelliGame}}\xspace}
	\newcommand{\IntelliJ}{\textsc{\mbox{IntelliJ}}\xspace}
	
	\begin{abstract}
		
		Interruptions and context switches resulting from meetings, urgent tasks, emails, and queries from colleagues contribute to productivity losses in developers' daily routines. This is particularly challenging for tasks like software testing, which are already perceived as less enjoyable, prompting developers to seek distractions. To mitigate this, applying gamification to testing activities can enhance motivation for test writing. One such gamification tool is \gamekins, which integrates challenges, quests, achievements, and leaderboards into the \Jenkins CI (continuous integration) platform. However, as \gamekins is typically accessed through a browser, it introduces a context switch. This paper presents an \emph{IntelliJ plugin} designed to seamlessly integrate \gamekins' gamification elements into the IDE, aiming to minimize context switches and boost developer motivation for test writing.
		
	\end{abstract}

	\ccsdesc[500]{Software and its engineering~Software testing and debugging}
	\ccsdesc[500]{Software and its engineering~Integrated and visual development environments}
	
	\keywords{Gamification, IDE, IntelliJ, Software Testing, Continuous Integration}
	
	\maketitle
	
	\section{Introduction}
	
	\gamekins~\cite{DBLP:conf/icse/StraubingerF22} is a tool that integrates gamification into the software testing process within the \Jenkins continuous integration platform, to encourage developers to improve their testing practices. Addressing the challenge of achieving high software quality and recognizing the common lack of motivation for testing among developers, \gamekins incorporates gamification elements like challenges, quests, leaderboards, and achievements. Developers earn points by completing test-related challenges and quests, engaging in leaderboard competition, and seeking achievements, with the integration into \Jenkins aiming to make gamification easily accessible to developers without additional training or resources.
	
	In a recent study~\cite{DBLP:conf/icse/StraubingerF23} we integrated \gamekins into an undergraduate software testing course to motivate students to engage more with testing. This study revealed a positive correlation between students' testing behavior and the use of \gamekins, resulting in a significant improvement in correct results compared to a non-gamified cohort. Additionally, students expressed satisfaction with the tool, confirming its effectiveness as a teaching tool.
	
	However, a notable issue reported by the students in this experiment was the necessity to switch from their Integrated Development Environment (IDE) to a browser with \gamekins open, causing a productivity dip and time loss navigating to \Jenkins. Such context switches, which represent a known productivity challenge~\cite{DBLP:conf/sigsoft/MeyerFMZ14}, are a concern not only for students but also for developers, who spend approximately 50\% of their work time in the IDE~\cite{DBLP:journals/tse/MeyerBBZ21}.
	
	In other work, we established that gamification of testing is possible directly in the IDE. Our \IntelliGame plugin for the \IntelliJ Java IDE\footnote{\url{https://www.jetbrains.com/idea/}}
	implements a multi-level achievement system rewarding positive testing behavior~\cite{DBLP:conf/icse/Straubinger024}. In a controlled experiment with 49 participants  \IntelliGame demonstrated clear improvements in testing behavior, for example, causing users to write more tests, achieve higher coverage and mutation scores, run tests more frequently, and achieve functionality earlier.
	
	In this paper, we aim to use the experience of gamifying testing directly in the IDE to address the problem of context switches caused by gamification in the CI. We therefore developed a new plugin for the \IntelliJ Java IDE that seamlessly integrates the functionality of \gamekins directly into the IDE, offering enhanced features to improve developers' workflow.

	\section{Gamification Elements of Gamekins}
	
	\subsection{Challenges}
	
	\gamekins offers a variety of different challenges, which are test and quality-oriented tasks for the developer to solve. There are currently eight different types of challenges available\footnote{Detailed information can be found in prior work~\cite{DBLP:conf/icse/StraubingerF22,DBLP:conf/icse/StraubingerF23}}:
	
	\begin{itemize}
		\item \textbf{Build Challenge}: This challenge shows the developer that the build on the CI failed and has to be fixed.
		\item \textbf{Test Challenge}: This generic challenge shows that the developer has to write a new test.
		\item \textbf{Class Coverage Challenge}: This challenge tasks the developer to cover more lines in a specific class.
		\item \textbf{Method Coverage Challenge}: This challenge focuses on improving the coverage of a specific method.
		\item \textbf{Line Coverage Challenge}: This challenge assigns the developer an uncovered line that they have to cover.
		\item \textbf{Branch Coverage Challenge}: This challenge focuses on the improvement of branch coverage of a covered line.
		\item \textbf{Mutation Challenge}: This challenge tasks the developer to detect a mutant generated by PIT\footnote{\url{https://pitest.org/}} by writing a new test.
		\item \textbf{Smell Challenge}: This challenge focuses on the removal of code and test smell with the help of SonarLint\footnote{\url{https://www.sonarsource.com/products/sonarlint/}}.
	\end{itemize}
	
	\subsection{Quests}
	
	Since the original inception as described in prior work~\cite{DBLP:conf/icse/StraubingerF22,DBLP:conf/icse/StraubingerF23}, the way quests are implemented has changed. Instead of having multiple challenges solved after each other, quests are now test- and quality-related tasks that involve making multiple quality improvements and interacting with gamification elements. It is therefore not bound to challenges anymore, but rather on improving different test and quality-related metrics. The following types of quests are available:
	
	\begin{itemize}
		\item \textbf{Add Tests Quest}: This quest tasks the developer to add a specified number of tests to the existing test suite.
		\item \textbf{Cover Branches Quest}: This quest focuses on covering an additional specified number of branches with new tests.
		\item \textbf{Cover Lines Quest}: This quest focuses on covering an additional specified number of lines with new tests.
		\item \textbf{Receive Challenges Quest}: This quest promotes interaction with other developers by receiving a challenge from another developer.
		\item \textbf{Send Challenges Quest}: This quest promotes interaction with other developers by sending a challenge to them.
		\item \textbf{Solve Achievements Quest}: This quest focuses on the completion of a new achievement.
		\item \textbf{Solve Challenges Quest}: This quest tasks the developer to solve a specified number of new challenges of one type.
		\item \textbf{Solve Challenges Without Rejection Quest}: This quest focuses on solving a specified number of challenges regardless of their types without rejecting one in between.
	\end{itemize}
	
	\subsection{Achievements}
	
	Developers are rewarded based on their testing accomplishments, encompassing various achievements of differing difficulty levels. These achievements range from simple tasks like adding a test to a project to more challenging objectives such as achieving 100\% coverage. Certain achievements remain hidden until they are completed and are tied to individual actions, like the addition of new tests. Achievements are triggered when users commit to the project repository. 
	There are two main types of achievements: individual achievements, which involve completing a specific number of challenges, and project-level achievements, which are linked to reaching a particular threshold within the project. 
	
	\section{IntelliJ plugin for Gamekins}

	The \toolname can be installed on any \IntelliJ Community Edition or Ultimate IDE, adding a new tab on the right side of the IDE interface. Upon initial access to this tab, a login prompt will appear, prompting users to enter their username, password, URL, and project name to establish a connection with the relevant Jenkins instance. Subsequent logins are streamlined through the use of a token, eliminating the need for repeated manual logins, although users have the flexibility to log out as needed. Project and user configurations are centrally managed within \gamekins on a \Jenkins server rather than within the \toolname. This allows for extensive collaboration within the project, and the analyses conducted by \gamekins need not be executed locally. A customized API integrated into \gamekins empowers the plugin to fetch challenges, quests, and user data from \gamekins, as well as transmit data to it. The \toolname tab encompasses various pages presenting all the information and gamification elements generated and provided by \gamekins, along with the buttons on the bottom to change between these pages (\cref{fig:challenges}).
	
	\subsection{Challenges}
	
	\begin{figure*}
		\includegraphics[width=\linewidth]{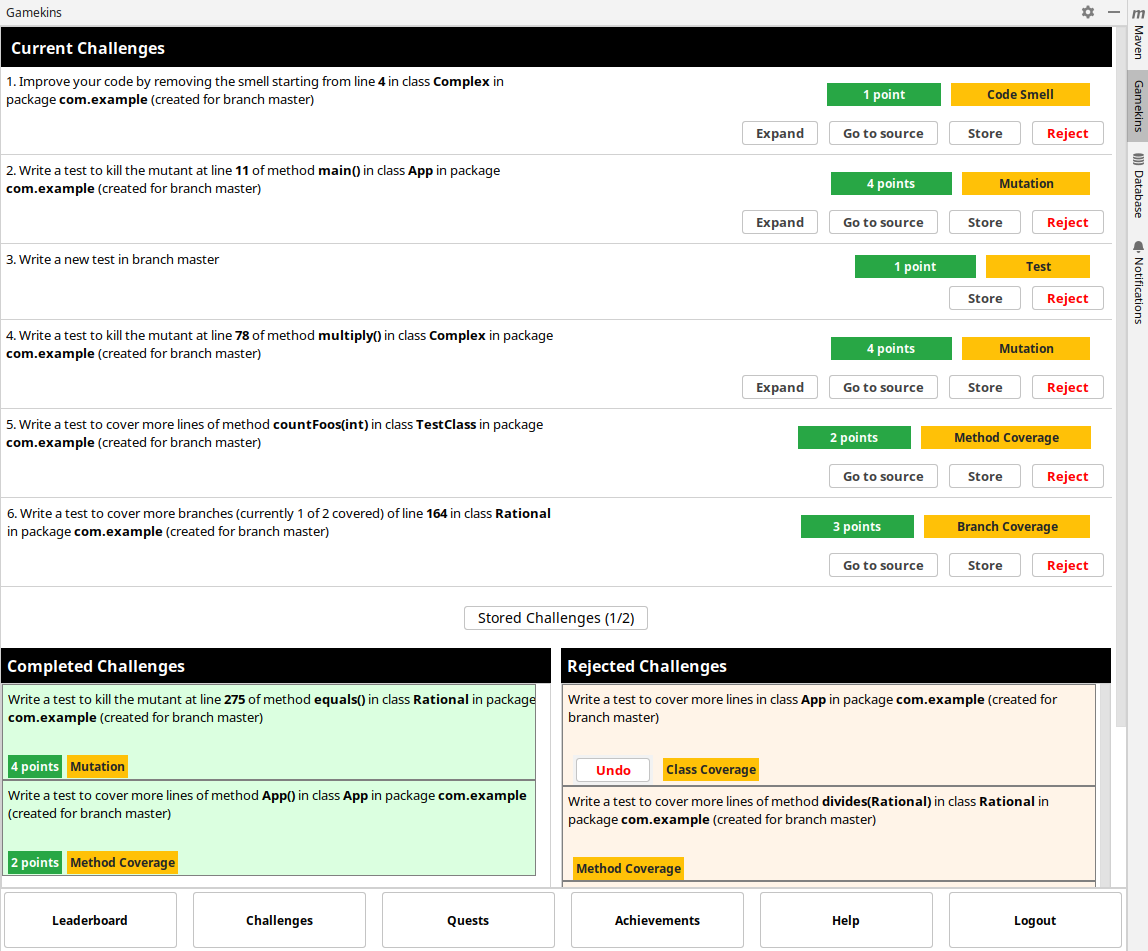}
		\vspace{-1em}
		\caption{Current, completed and rejected challenges}
		\label{fig:challenges}
	\end{figure*}
	
	All challenges are visible in the \toolname, with the availability of different information and actions based on the challenge type. For instance, a Test Challenge solely displays information about the challenge without further interactions with \IntelliJ (\cref{fig:challenges}). Similarly, a Build Challenge provides information and includes an additional link to the log files in \Jenkins. The Class and Method Coverage Challenges offer a button to navigate to and highlight the respective class or method. This feature is also available for Line and Branch Coverage challenges, highlighting the specified line. The Branch Coverage Challenge additionally provides information about covered and uncovered branches for the tested line. The Mutation and Smell Challenges allow developers to go to and highlight the original line of code or the start of the test smell, respectively. These challenges also have an expandable area containing details such as the mutant or a smell description.
	
	\begin{figure}
		\includegraphics[width=\linewidth]{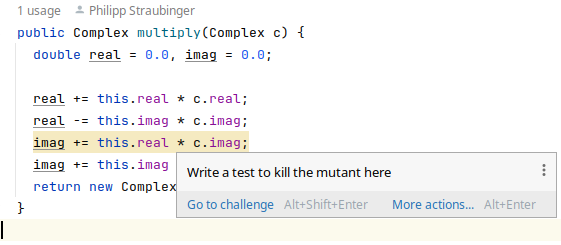}
		\vspace{-1em}
		\caption{Highlighted line of code with the tooltip giving information about the challenge}
		\label{fig:highlight}
	\end{figure}
	
	The challenges are not only visible in the \toolname tab but also feature yellow highlighting in the source files, aiding developers in easily identifying challenges in their current code. Hovering over the highlighted code snippet reveals a tooltip with the challenge description and a button to open the challenges page in the \toolname (\cref{fig:highlight}). This page displays current challenges along with completed and rejected ones (\cref{fig:challenges}).
	
	Developers can reject challenges with an explanation if they deem them irrelevant, prompting the generation of new challenges. Once rejected, challenges cannot be reinstated, except for Class Coverage Challenges, which can be restored using the Undo button (\cref{fig:challenges}). If a developer finds a challenge interesting but unsuitable for the current workflow, they can store a configurable number of challenges. These stored challenges can also be shared with a colleague working on the same project, especially if the source code is segmented into specialized fields unknown to the participant.
	
	\begin{figure}
		\includegraphics[width=\linewidth]{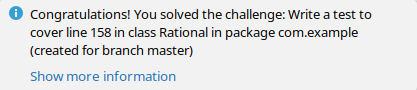}
		\vspace{-1em}
		\caption{Notification received after solving a challenge}
		\label{fig:notification}
	\end{figure}
	
	To address a challenge, developers simply need to commit and push their changes to their version control system, triggering \Jenkins to run the CI pipeline in the background. Developers can seamlessly continue their work in the project as \gamekins automatically executes after the pipeline concludes, providing notifications upon completion. General notifications about the finished build and specific notifications for each solved and generated challenge or quest are displayed (\cref{fig:notification}) in \IntelliJ. These notifications include links to view the corresponding elements in the \toolname. \vspace{2cm}
	
	\subsection{Quests}
	
	\begin{figure}
		\includegraphics[width=\linewidth]{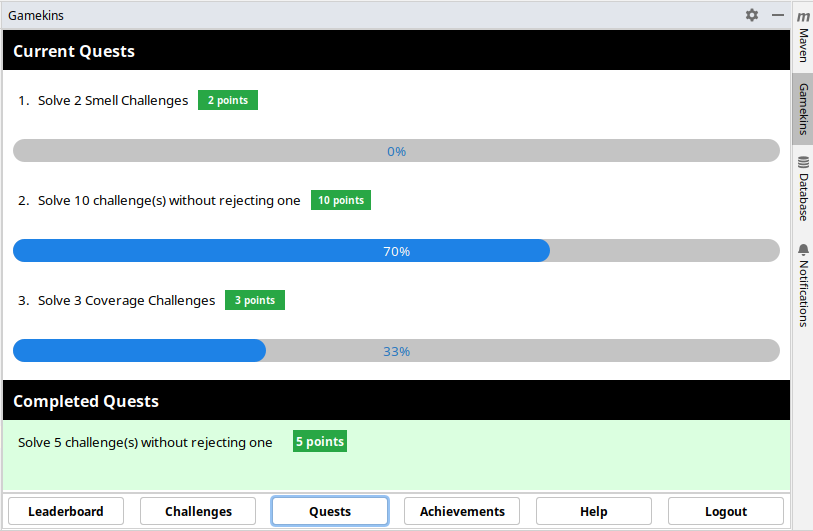}
		\vspace{-1em}
		\caption{Current and completed quests}
		\label{fig:quests}
	\end{figure}
	
	Each quest in the \toolname is displayed on the quests page (\cref{fig:quests}) along with its description and a progress bar. Both ongoing and completed quests are shown. The progress bar indicates the percentage of quest completion and is dynamically updated whenever relevant actions are undertaken.
	
	\subsection{Achievements}
	
\begin{figure}
		\includegraphics[width=\linewidth]{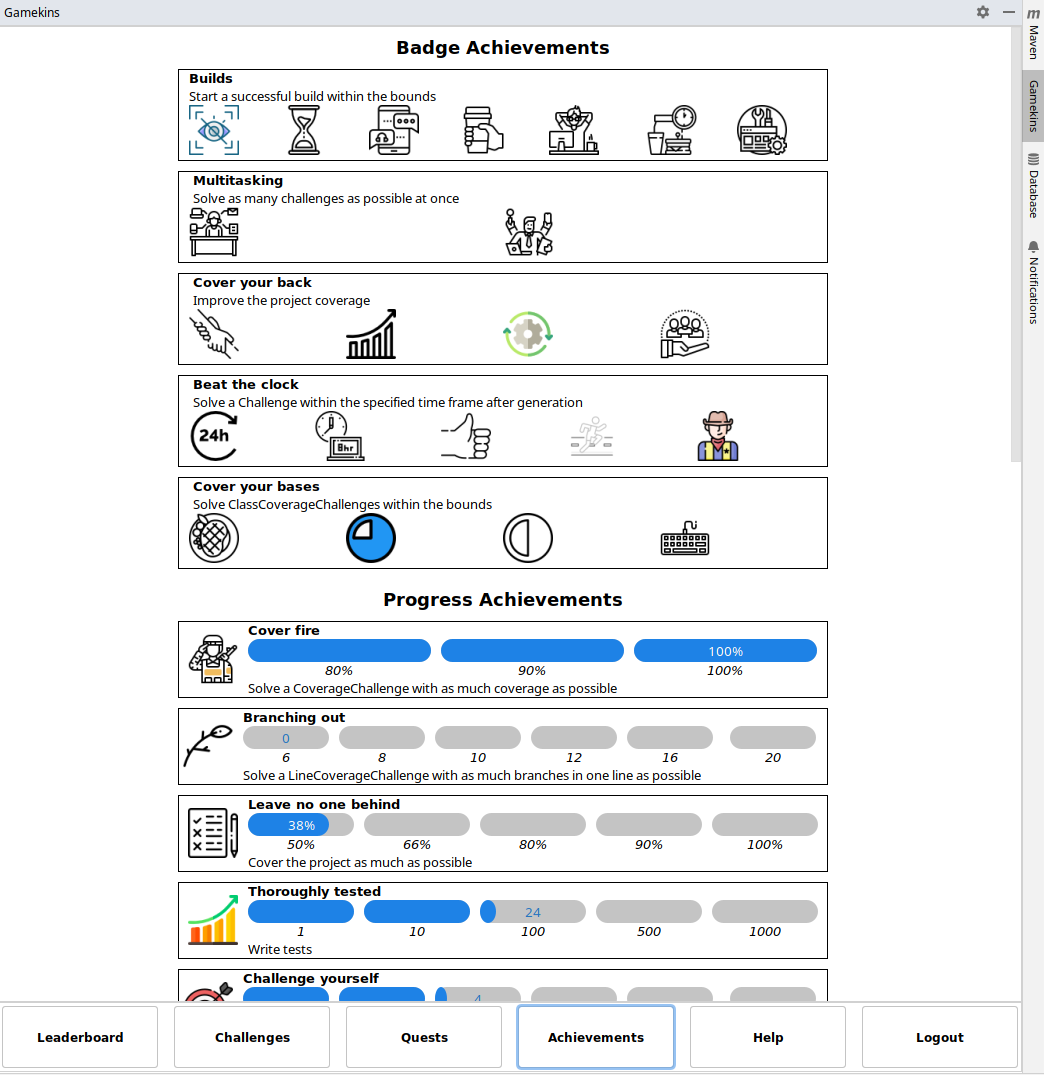}
		\vspace{-1em}
		\caption{List of achievements}
		\label{fig:achievements}
	\end{figure}

	Presently, the \toolname encompasses all achievements from \gamekins, allowing developers to access a dedicated tab to view both acquired and available achievements (\cref{fig:achievements}). Each achievement consists of an icon, a title, a description, and the date and time it was awarded.
	
	\subsection{Leaderboards}
	
	\begin{figure}
		\includegraphics[width=\linewidth]{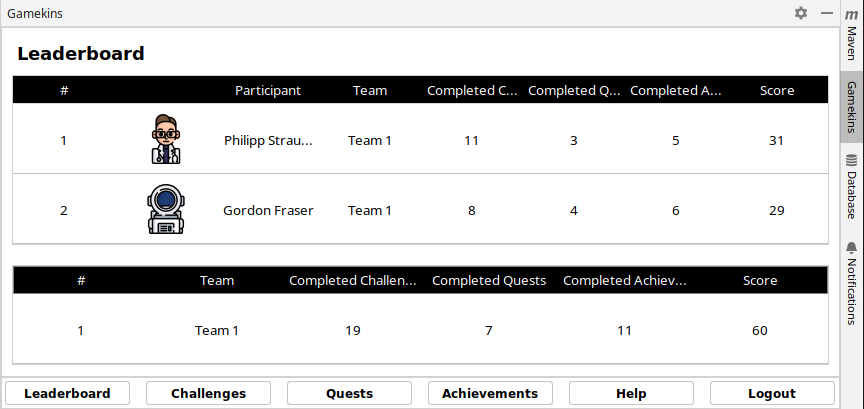}
		\vspace{-1em}
		\caption{Example leaderboard}
		\label{fig:leaderboard}
	\end{figure}
	
	To incentivize developers to solve challenges and complete quests, points are awarded based on the task's difficulty. These points are accumulated for each user and team, prominently displayed in a dedicated leaderboard within the \toolname tab (\cref{fig:leaderboard}). The leaderboard showcases rankings for individual users as well as teams, emphasizing their scores in the logged-in project. Information regarding completed challenges, quests, achievements, and earned points is presented on each leaderboard. Additionally, users have the option to personalize their appearance by selecting one of 50 avatars available within \gamekins.
	
	\section{Conclusions}
	
	The \toolname seamlessly incorporates the functionalities of \gamekins into the IDE, presenting essential information such as leaderboards, challenges, quests, and achievements directly within the developer's view. This integration minimizes context switches and associated productivity losses. We are currently enhancing the achievements system of \gamekins and introducing new ones. These improvements, along with future enhancements, will be integrated into \toolname. Additionally, we plan to assess the plugin in our university courses to evaluate its effectiveness. Lastly, we aim to introduce new types of challenges and achievements, building upon our experience with \IntelliGame.
	
	\noindent\toolname is available at:
	\begin{center}
		\href{https://github.com/se2p/Gamekins-IntelliJ-Plugin}{https://github.com/se2p/Gamekins-IntelliJ-Plugin}
	\end{center}
	
	\section{Acknowledgements}
	This work is supported by the DFG under grant FR 2955/2-1
	
	\balance
	\bibliographystyle{ACM-Reference-Format}
	\bibliography{bib}
	
\end{document}